\documentclass[12pt]{article}
\usepackage{amssymb}
\usepackage{amsmath}
\usepackage{amsfonts}
\usepackage{epsfig}
\usepackage{anysize}
\marginsize{2cm}{2cm}{2.5cm}{3cm}
\date{}

\begin{document}

\title{Incidence of $q$-statistics in rank distributions}

\author{G. Cigdem Yalcin\textsuperscript{1}, Alberto Robledo\textsuperscript{2}, Murray Gell-Mann\textsuperscript{3}\\
\footnotesize 1. Department of Physics, Istanbul University, 34134, Vezneciler, Istanbul, Turkey.\\
\footnotesize 2. Instituto de Fisica y Centro de Ciencias de la Complejidad, Universidad Nacional Autonoma de Mexico, \\
\footnotesize Apartado Postal 20-364, M\'{e}xico 01000 D.F., Mexico.\\
\footnotesize 3. Santa Fe Institute, Santa Fe, NM 87501.\\
       }




\maketitle

\begin{abstract}
{We show that size-rank distributions with power-law decay (often only over a
limited extent) observed in a vast number of instances in a widespread
family of systems obey Tsallis statistics. The theoretical framework for
these distributions is analogous to that of a nonlinear iterated map near a
tangent bifurcation for which the Lyapunov exponent is negligible or
vanishes. The relevant statistical-mechanical expressions associated with
these distributions are derived from a maximum entropy principle with the
use of two different constraints, and the resulting duality of entropy
indexes is seen to portray physically relevant information. While the value
of the index $\alpha $ fixes the distribution's power-law exponent, that for
the dual index $2-\alpha $ ensures the extensivity of the deformed entropy.}
\end{abstract}

\indent \ \  rank-ordered data |  generalized entropies

\vspace*{10pt}

Zipf's law refers to the (approximate) power law obeyed by sets of data when
these are sorted out and displayed by rank in relation to magnitude or rate
of recurrence \cite{general1}. The sets of data originate from many
different fields: astrophysical, geophysical, ecological, biological,
technological, financial, urban, social, etc., suggesting some kind of
universality. Over the years this circumstance has attracted much attention
and the rationalization of this empirical law has become a common endeavor
in the study of complex systems \cite{schroeder, gell-mann}. Here we
pursue further the view \cite{robledo1,robledo2} that an understanding of
the omnipresence of this type of rank distribution hints to an underlying
structure similar to that which confers systems with many degrees of freedom
the familiar macroscopic properties described by thermodynamics. That is,
the quantities employed in describing this empirical law obey expressions
derived from principles akin to a statistical-mechanical formalism \cite%
{robledo1,robledo2}. The most salient result presented here is that the
reproduction of the data via a maximum entropy principle indicates that
access to its configurational space is severely hindered to a point that the
allowed configurational space has a vanishing measure. This feature appears
to be responsible for the entropy expression not to be of the
Boltzmann-Gibbs or Shannon type but instead it takes that of the Tsallis
form \cite{tsallis1}, while the extensivity of entropy is preserved. It is perhaps worth
clarifying that our study is set in discrete space and it does not consider any formal
Hamiltonian system.

In Fig. 1 we show three examples of ranked data that appear to display
power-law behavior along a considerable large interval of rank values. In
the top panels of this figure we show data for the wealth of billionaires in
the US \cite{forbes1}, in the middle panels data for the energy released by
earthquakes in California \cite{earthquakes1}, and in the bottom panels data
for the intensity of solar flares \cite{solarflares1}. In the left panels
logarithmic scales are used for both size and rank, whereas the right panels
show the same data in log-linear scales. The left panels indicate
approximate power law decay for large rank and a clear deviation from this
for small to moderate rank. As we shall show below the theoretical
description reproduces the data in Fig. 1 for the entire rank interval.

In Section II we recall \cite{robledo1,robledo2, pietronero1} the concise stochastic
approach for raw data generated by a power-law distribution $P$ for the size
random variable $N$ that yields an analytical expression for the size-rank
distribution $N(k)$. This analytical expression involves a deformed
exponential that has been shown to reproduce quantitavely real data and has
as a limiting form the classical Zipf law \cite{robledo1,robledo2, pietronero1}. We also
recall \cite{robledo1,robledo2} the analogy that exists between the
stochastic approach and the deterministic nonlinear dynamics at and close to
the tangent bifurcation. This analogy allows for a convenient description of
finite-sized data that deviates from power-law behavior for both small and
large rank. In Section III we derive the rank distribution $N(k)$ from a
maximum entropy principle (MEP) and this allows us, via a well-known
deformation index duality, to discuss two different entropy expressions of
the Tsallis type obtained from two different sets of constraints \cite%
{thurner hanel gell-mann1, thurner hanel gell-mann2,thurner hanel gell-mann3}.
The values of the two entropy expressions
coincide but they yield different information for the set of data under
consideration. In Section IV we use this duality to discuss entropy
extensivity of the ranked data and the presence of a strong phase-space
contraction. This is shown to be the source of a generalized entropy that
departs from the usual Shannon expression. This departure is extreme for the
classical Zipf case, implying that the data can sample only a set of zero
measure. Finally, in Section V we discuss and summarize our results.


\subsection*{Significance.}
The contents presented are of prime importance to the field of generalized statistical mechanics. We fulfill a longstanding need of exhibiting the kind of abundant real world data that matches the formal developments in this subject. These are size-rank distributions for which we provide a solid bridge between experimental data and theory. Also, this work delivers a working explanation for the existing duality between the two Tsallis-type entropy expressions that generalize the canonical expression. One relates to the distribution power-law exponent whereas the other ensures entropy extensivity. The generalized entropies arise from a drastic reduction of configurations available to the system. We argue that this phase-space contraction is farthest for ranked data of the Zipf type.



\section{The distribution functions that generate Zipf's law}

A basic approach for the study of ranked data consists of three simply
related distribution functions \cite{robledo1,robledo2, pietronero1}. The
input is the distribution $P(N)$ of the data $N$ under consideration, that
is, it is assumed that the data is generated by a source described by $P(N)$
such that $N$ can be thought of as a random variable. With no loss of
generality we restrict $N$ to take positive values within an interval $%
N_{\min }\leq N\leq N_{\max }$, where we allow for the limiting
possibilities $N_{\min }=0$\ and/or $N_{\max }\rightarrow \infty $. The
total number of data extracted from $P(N)$ is denoted by $\mathcal{N}$.
Next, the (complementary) cummulative distribution $\Pi (N,N_{\max })$ is
determined from $P(N)$,

\begin{equation}
\Pi (N,N_{\max })=\int\limits_{N}^{N_{\max }}P(N^{\prime })dN^{\prime },
\label{cummulative}
\end{equation}%
where the normalization of $P(N)$ implies $\Pi (N_{\min },N_{\max })=1$. We
can recover $P(N)$ from $\Pi (N,N_{\max })$,%
\begin{equation}
P(N)=-\frac{\partial }{\partial N}\Pi (N,N_{\max }).  \label{basic2}
\end{equation}%
By construction, the distribution $\Pi (N,N_{\max })$ sorts out data
according to its magnitude: As $N$ is decreased from $N_{\max }$ the
distribution $\Pi $ increases monotonically taking values from $\Pi (N_{\max
},N_{\max })=0$ to $\Pi (N_{\min },N_{\max })=1$, so it can be identified
with $k/\mathcal{N}$, where $k$ is the rank and $\mathcal{N}$ is the total
number of data extracted from $P(N)$, and $k_{\max }=\mathcal{N}$. The last
and third distribution is the size-rank function $N(k)$ and can be obtained
by solving

\begin{equation}
\frac{k}{\mathcal{N}}=\int\limits_{N(k)}^{N_{\max }}P(N^{\prime })dN^{\prime
},  \label{cummulative2}
\end{equation}%
for $N(k)$. If $k$ is to be an integer the possible lower limits in the
integral in Eq. (\ref{cummulative2}), $N(1)$, $N(2)$, $...$, $N(k_{\max })$
are such that the integral takes values $1/\mathcal{N}$, $2/\mathcal{N}$, $%
...$, $k_{\max }/\mathcal{N}$.

If we make use of a power law form for $P(N)$,

\begin{equation}
P(N)\sim N^{-\alpha },\ 1\leq \alpha <\infty ,  \label{basic}
\end{equation}%
we have \cite{pietronero1,robledo1,robledo2}

\begin{eqnarray}
\Pi (N(k),N_{\max }) &=&\int\limits_{N(k)}^{N_{\max }}N^{-\alpha }dN  \notag
\\
&=&\frac{1}{1-\alpha }\left[ N_{\max }^{1-\alpha }-N(k)^{1-\alpha }\right] ,
\label{cummulative4}
\end{eqnarray}%
or, in \ terms of the $q$-deformed logarithmic function $\ln _{q}(x)\equiv
(1-q)^{-1}[x^{1-q}-1]$ with $q$ a real number,

\begin{equation}
\ln _{\alpha }N(k)=\ln _{\alpha }N_{\max }-\mathcal{N}^{-1}k.
\label{logalpha1}
\end{equation}%
The size-rank distribution $N(k)$ is explicitly obtained from the above with
use of the inverse of $\ln _{q}(x)$, the $q$-deformed exponential function $%
\exp _{q}(x)\equiv \left[ 1+(1-q)x\right] ^{1/(1-q)}$, this is%
\begin{equation}
N(k)=N_{\max }\exp _{\alpha }(-N_{\max }^{\alpha -1}\mathcal{N}^{-1}k).
\label{size-rank1}
\end{equation}%
When $\alpha =1$ Eq. (\ref{size-rank1}) acquires the ordinary exponential
form

\begin{equation}
N(k)=N_{\max }\exp (-\mathcal{N}^{-1}k),  \label{size-rank2}
\end{equation}%
whereas in the limit $N_{\max }\rightarrow \infty $ Eq. (\ref{size-rank1})
becomes the power law $N(k)\sim k^{1/(1-\alpha )}$ that when $\alpha =2$
gives the simple classical Zipf's law form $N(k)\sim k^{-1}$.

An explicit analogy between the generalized law of Zipf and the nonlinear
dynamics of intermittency has been studied \cite{robledo1, robledo2}. We
recall the renormalization group (RG) fixed-point map for the tangent bifurcation. The trajectories $%
x_{t},\ t=1,2,3,...$, produced by this map, comply (analytically) with:%
\begin{equation}
\ln _{z}x_{t}=\ln _{z}x_{0}+ut  \label{trajectory2}
\end{equation}%
or 
\begin{equation}
x_{t}=x_{0}\exp _{z}\left[ x_{0}^{z-1}ut\right]   \label{trajectory3}
\end{equation}%
where the $x_{0}$ are the initial positions. The parallels between Eqs. (\ref%
{trajectory2}) and (\ref{trajectory3}) with Eqs. (\ref{logalpha1}) and (\ref%
{size-rank1}), respectively, is clear, and therefore, we conclude that the
dynamical system represented by the fixed-point map operates in accordance
to the same $q$-generalized statistical-mechanical properties discussed
below. We notice that the absence of an upper bound for the rank $k$ in Eqs.
(\ref{logalpha1}) and (\ref{size-rank1}) is equivalent to the tangency
condition in the map. Accordingly, to describe data with finite maximum
rank, we look at the changes in $N(k)$ brought about by shifting the
corresponding map from tangency, \textit{i.e.}, we consider the
trajectories, $x_{t}$, with initial positions $x_{0}$ of the map:%
\begin{equation}
x^{\prime }=x\exp _{z}(ux^{z-1})+\varepsilon ,\;0<\varepsilon \ll 1
\label{offtangency1}
\end{equation}%
with the identifications $k=t$, $\mathcal{N}^{-1}=-u$, $N(k)=x_{t}+x^{\ast }$%
, $N_{\max }=x_{0}+x^{\ast }$ and $\alpha =z$, where the translation, $%
x^{\ast }$, ensures that all $N(k)\geq 0$. The capability of this approach
to reproduce quantitatively real data for ranked data with deviations
from power law for large rank has been discussed \cite{robledo1, robledo2}.

\section{Rank distributions from maximum entropy principle}

The rank distribution $N(k)$ described in the previous section can be
obtained from a maximum entropy principle (MEP), and, as we shall see, this
allows to put forward important interpretations regarding the nature of the
systems that give rise to it. But first we adjust our interpretation of $N(k)
$. This quantity is actually the size or magnitude of the data under
consideration, the number of units that, in a microcanical ensemble
description, is the number of configurations that take place for a fixed
value of $k$. Therefore its inverse, $p_{k}=1/N(k)$, is the (uniform)
probability for the occurrence of each unit that constitutes $N(k)$. The
probability $p_{k}$ is normalized for fixed $k$, and we denote its limiting
values by $p_{\min }=1/N_{\max }$ and $p_{\max }=1/N_{\min }$, $N_{\min
}\leq N(k)\leq N_{\max }$.

A formal investigation of the possible entropy expressions that generalize the Boltzmann-Gibbs
or Shannon canonical form has been systematically carried out with the use of the MEP under the assumption that only three of the Shannon-Kinchin axioms
hold \cite{thurner hanel gell-mann1, thurner hanel gell-mann2, thurner hanel gell-mann3}.
(Inclusion of the fourth, composability, uniquely defines the canonical form). Here we focus only
on the Tsallis expressions \cite{abe1}. 

Consider the entropy functional $\Phi _{1}[p_{k}]$ with Lagrange multipliers 
$a$ and $b$,%
\begin{equation}
\Phi _{1}[p_{k}]=S_{1}[p_{k}]+a\left[ \sum_{k=0}^{k_{\max }}p_{k}-\mathcal{P}%
\right] +b\left[ \sum_{k=0}^{k_{\max }}kp_{k}-\mathcal{K}\right] ,
\label{functional1}
\end{equation}%
where the entropy expression $S_{1}[p_{k}]$ has the trace form \cite{thurner
hanel gell-mann1}%
\begin{equation}
S_{1}[p_{k}]=\sum_{k=0}^{k_{\max }}s_{1}(p_{k}).  \label{entropy1}
\end{equation}%
Optimization via $\partial \Phi _{1}[p_{k}]/\partial p_{k}=0$, $%
k=0,1,2,...,k_{\max }$, gives 
\begin{equation}
s_{1}^{\prime }(p_{k})=-a-bk.  \label{derentropy1}
\end{equation}%
Now, the choices%
\begin{equation}
s_{1}^{\prime }(p_{k})=\alpha \ln _{\alpha }p_{k}^{-1}-1,\;a=-\alpha \ln
_{\alpha }p_{\min }^{-1}+1,\;b=\alpha \mathcal{N}^{-1},  \label{choices1}
\end{equation}%
lead to%
\begin{equation}
\ln _{\alpha }\ p_{k}^{-1}=\ln _{\alpha }p_{\min }^{-1}-\mathcal{N}^{-1}k.
\label{logalphapk1}
\end{equation}%
or%
\begin{equation}
p_{k}^{-1}=p_{\min }^{-1}\exp _{\alpha }(-p_{\min }^{1-\alpha }\mathcal{N}%
^{-1}k).  \label{expalphapk1}
\end{equation}%
from which we immediately recover Eqs. (\ref{logalpha1}) and (\ref%
{size-rank1}).

We repeat the same optimization procedure but with a constraint change \cite%
{thurner hanel gell-mann1}. Consider the functional $\Phi _{2}[p_{k}]$ with
Lagrange multipliers $c$ and $d$,%
\begin{equation}
\Phi _{2}[p_{k}]=S_{2}[p_{k}]+c\left[ \sum_{k=0}^{k_{\max }}p_{k}-\mathcal{P}%
\right] +d\left[ \sum_{k=0}^{k_{\max }}kp_{k}^{\alpha ^{\prime }}-\mathcal{K}%
_{\alpha ^{\prime }}\right] ,  \label{functional2}
\end{equation}%
and where the entropy expression $S_{2}[p_{k}]$ has also a trace form

\begin{equation}
S_{2}[p_{k}]=\sum_{k=0}^{k_{\max }}s_{2}(p_{k}).  \label{entropy2}
\end{equation}%
Optimization via $\partial \Phi _{2}[p_{k}]/\partial p_{k}=0$, $%
k=0,1,2,...,k_{\max }$, gives

\begin{equation}
s_{2}^{\prime }(p_{k})=-c-dk.  \label{derentropy2}
\end{equation}%
And this time the choices

\begin{equation}
s_{2}^{\prime }(p_{k})=-(2-\alpha ^{\prime })\ln _{\alpha ^{\prime
}}p_{k}-1,\;c=(2-\alpha ^{\prime })\ln _{\alpha ^{\prime }}p_{\min
}+1,\\\;
  \end{equation}
  \begin{equation}
d=(2-\alpha ^{\prime })\mathcal{N}^{-1},  \label{choices2}
\end{equation}%

give the expressions

\begin{equation}
\ln _{\alpha 
{\acute{}}%
}\ p_{k}=\ln _{\alpha 
{\acute{}}%
}\ p_{\min }+\mathcal{N}^{-1}k.  \label{logalphaprimepk1}
\end{equation}%
or%
\begin{equation}
p_{k}=p_{\min }\exp _{\alpha 
{\acute{}}%
}(p_{\min }^{\alpha 
{\acute{}}%
-1}\mathcal{N}^{-1}k).  \label{expalphaprimepk1}
\end{equation}

A comparison of Eqs. (\ref{logalpha1}) and (\ref{size-rank1}) with Eqs. (\ref%
{logalphaprimepk1}) and (\ref{expalphaprimepk1}), respectively, indicates
that they become equivalent with the identifications%
\begin{equation*}
p_{k}=1/N(k),\;p_{\min }=1/N_{\max },\;\alpha ^{\prime }=2-\alpha .
\end{equation*}%
Furthermore, $s_{2}^{\prime }(p_{k})=s_{1}^{\prime }(p_{k})$ (as given by
Eqs. (\ref{derentropy1}), (\ref{choices1}), (\ref{derentropy2}) and (\ref%
{choices2})) and therefore%
\begin{equation}
S_{2}[p_{k}]=S_{1}[p_{k}],  \label{entropyequality1}
\end{equation}%
where their optimized expressions are%
\begin{equation}
S_{1}[p_{k}]=\sum_{k=0}^{k_{\max }}p_{k}\ln _{\alpha }\ p_{k}^{-1},
\label{entropy1opt}
\end{equation}%
and%
\begin{equation}
S_{2}[p_{k}]=-\sum_{k=0}^{k_{\max }}p_{k}\ln _{\alpha ^{\prime }}p_{k}.
\label{entropy2opt}
\end{equation}

Under the assumption of validity of only the first three Shannon-Kinchin axioms
it has been shown \cite{thurner hanel gell-mann1, thurner hanel gell-mann2}
that there are only two ways to construct entropy expressions via the MEP procedure.
These correspond to the constraints used in Eqs. (\ref{functional1}) and (\ref%
{functional2}) and the resulting entropy expressions are those in Eqs. (\ref%
{entropy1opt}) and (\ref{entropy2opt}). The two approaches are related via
the deformation index duality $\alpha ^{\prime }=2-\alpha $, and, for the
same distribution $p_{k}$, their values are equal as in Eq. (\ref%
{entropyequality1}). For an earlier account of this duality property see
Ref. \cite{robledo3}. See also \cite{robledo4}. From our earlier discussion we know
that the index $\alpha $ fixes the shape of the rank distribution $N(k)$ and
that its departure from unity generates its power-law feature and that the
value $\alpha =2$ reproduces the classic Zipf law. To complete the picture
we need to clarify the role of the dual index $\alpha ^{\prime }$ and the
distribution $p_{k}$, and from this obtain an understanding of the dual
entropy expressions in Eqs. (\ref{entropy1opt}) and (\ref{entropy2opt}).
Interestingly, when $\alpha =\alpha ^{\prime }=1$ the duality collapses into
the Boltzmann-Gibbs or Shannon entropy expressions and the exponential form
for $N(k)$, but for $\alpha =2$ we have $\alpha ^{\prime }=0$ and $p_{k}$
grows linearly with $k$.

In Fig. 2 we show the same three sets of data in Fig. 1 in log-linear
scales. This time we fit them with Eqs. (\ref{size-rank1}) and (\ref%
{expalphaprimepk1}) and observe that the data are well described with values
of the deformations $\alpha =2$ and $\alpha ^{\prime }=0$.

\section{Statistical mechanics of contracted configuration space}

The function $N(k)$ has the properties of a microcanonical partition
function \cite{robledo1, robledo2}. That is, the size $N(k)$ is the result of 
$N(k)$ equally-probable configurations, and the probabilities $p_{k}$ are
correspondingly normalized for fixed $k$. However, these probabilities are
not normalized if the rank $k$ runs across its values $k=0,...,k_{\max }$,
and we do not make an attempt here to do so. Instead, we look at the rank
dependence in Eq. (\ref {expalphaprimepk1}), that we identify as the system's size
dependence. As it can be observed in the right panels
of Fig. 2 the probabilities $p_{k}$ rises sharply and then saturates as $k$ increases. The pure deformed exponential 
\begin{equation}
\frac{p_{k}}{p_{\min }}=\frac{N_{\max}}{N(k)}=\exp _{\alpha 
{\acute{}}%
}(p_{\min }^{\alpha 
{\acute{}}%
-1}\mathcal{N}^{-1}k)\label{puredeformedexponential1}
\end{equation}%
measures the change in the number of microcanonical configurations with the size of the system $k$.
We define the size-dependent entropy%
\begin{equation}
S(k)\equiv\ln _{\alpha 
{\acute{}}%
}\left( \frac{N_{\max}}{N_{k}}\right) ,\;k\;\text{fixed}, \label{microcanonicalentropy1}
\end{equation}%
and from Eqs. (\ref{puredeformedexponential1}) and (\ref{microcanonicalentropy1}%
) we observe that $S(k)$ is extensive, doubling the numbers of billionaires,
earthquakes or solar flares in the data sets doubles the value of $S(k)$,
and it can be seen to be so because the deformation index $\alpha 
{\acute{}}%
$ has the precise value to ensure this property.
The constraint%
\begin{equation}
\sum_{k=0}^{k_{\max }}k\ p_{k}^{\alpha ^{\prime }}=\mathcal{K}_{\alpha
^{\prime }}  \label{constraint1}
\end{equation}%
in Eq. (\ref{functional2}) for entropy maximization indicates that the phase
space, $N_{\min }\leq N \leq N_{\max }$, is highly constrained since the
probabilities $p_{k}<1$ need to be enhanced up to $p_{k}^{\alpha ^{\prime }}$, $%
\alpha 
{\acute{}}%
<1$, in order to obtain a meaningful average of $k$. In relation to this,
notice that $N(k)$ is a monotonously decreasing function with a power law
feature. This phase-space contraction is extreme for the case of Zipf law
because $\alpha 
{\acute{}}%
$ reaches its minimum value of zero. For a system with normal occupation of
phase spase, the number of configurations grow exponentially and $S(k)$
above becomes extensive in $k$ for index value $\alpha 
{\acute{}}%
=1$ whereas the phase space in the most contracted stage the number of
configurations grow only linearly and and this linearity is preserved in $%
S(k)$ when $\alpha 
{\acute{}}%
=0$.   

In Fig. 3 we show the same data in Figs. 1 and 2 but this time plotted in
deformed logarithmic scales with deformation indexes $\alpha \simeq 2$ and $%
\alpha 
{\acute{}}%
\simeq 0$. Data in these scales are displayed linearly and should be fitted by the
theoretical expressions Eqs. (\ref{size-rank1}) and (\ref{expalphaprimepk1})
if these equations represent the behavior of the data.

\section{Discussion}

We have shown that size-rank distributions with power-law decay
for moderate and large values of rank obey Tsallis statistics. The
small-rank behavior that departs from the power law is also well reproduced
by the deformed exponential expression in Eq. (\ref{size-rank1}) for $N(k)$.
For the specific data we presented (US billionaires, California earthquakes
and solar-flare intensities) the values of the exponential deformations were
found to be $\alpha \simeq 2$, the value needed to obtain the classical Zipf
law. In order to advance further in the characterization of the apparent
relationship between rank distributions and generalized statistical
mechanics, such as that of Tsallis, we rederived Eq. (\ref{size-rank1}) for $%
N(k)$ from a maximum entropy procedure. This was done in accordance to the
consideration of validity of only the first three Shannon-Kinchin axioms \cite{thurner
hanel gell-mann1, thurner hanel gell-mann2}. Under these conditions duality
of entropy expressions appears according to
the use of two different constraints. In doing this we introduce the
(unormalized) distribution $p_{k}=1/N(k)$, actually $N(k)$ is the number of
data for the same rank $k$ (playing the role of a partition function) \cite%
{robledo1, robledo2}. We obtain equality of the entropy expressions $%
S_{1}[p_{k}]=S_{2}[p_{k}]$ in Eqs. (\ref{entropy1opt}) and (\ref{entropy2opt}%
) and a companion rank distribution expression for $p_{k}$, Eq. (\ref%
{expalphaprimepk1}). As it is known \cite{thurner hanel gell-mann1, thurner
hanel gell-mann2} the two entropies $S_{1}[p_{k}]$ and $S_{2}[p_{k}]$ correspond to the dual
deformation indexes $\alpha $ and $\alpha ^{\prime }=2-\alpha $. We have
enquired as to the different roles of the two entropy expressions and
identify the physically relevant information carried by each one. We found
that the value of the index $\alpha $ fixes the distribution's power-law
exponent for $N(k)$ and that the dual index $\alpha 
{\acute{}}%
=2-\alpha $ ensures the extensivity of the deformed entropy. Finally, we
argued that the value $\alpha =2$, that corresponds to the classical Zipf
law, manifests as $\alpha 
{\acute{}}%
=0$ that we interpret as an extreme contraction of the phase space from which
the data originates.

\subsection*{Aknowledgements}
G.C.Y and A.R. gratefully acknowledge the hospitality of the Santa Fe Institute.
Support by DGAPA-UNAM-IN100311 and CONACyT-CB-2011-167978 (Mexican Agencies) is acknowledged. G.C.Y. was supported by the Scientific Research Projects Coordination Unit of Istanbul University with project number 36529. M.G.-M. acknowledges the generous support of Insight Venture Partners and the Bryan J. and June B. Zwan Foundation.

\clearpage

\section*{Figures}

\vspace*{20pt}

\begin{figure*}[ht]
\begin{center}
\centerline{\includegraphics[width=.7\textwidth]{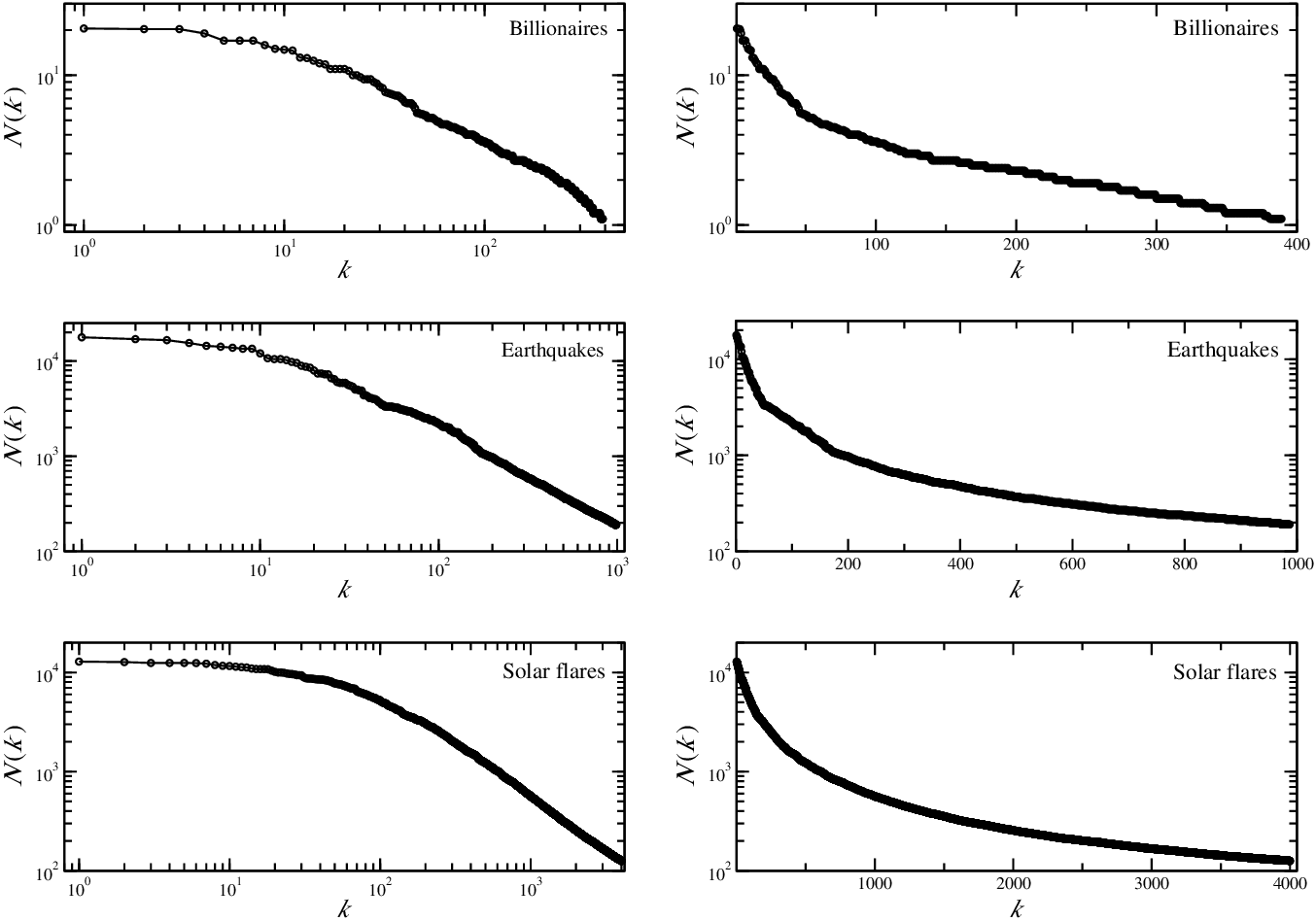}}
\caption{Three examples of ranked data that appear to display power-law
behavior along an interval of rank values. The top panels show data for
the wealth of billionaires in the US \protect\cite{forbes1}. The middle
panels present data for the energy released by earthquakes in California 
\protect\cite{earthquakes1}. The bottom panels provide data for the intensity
of solar flares \protect\cite{solarflares1}. In the left panels the data is
shown in logarithmic scales, whereas the right panels show the same data in
log-linear scales. See text for description.}\label{afoto}
\end{center}
\end{figure*}

\begin{figure*}[ht]
\begin{center}
\centerline{\includegraphics[width=.7\textwidth]{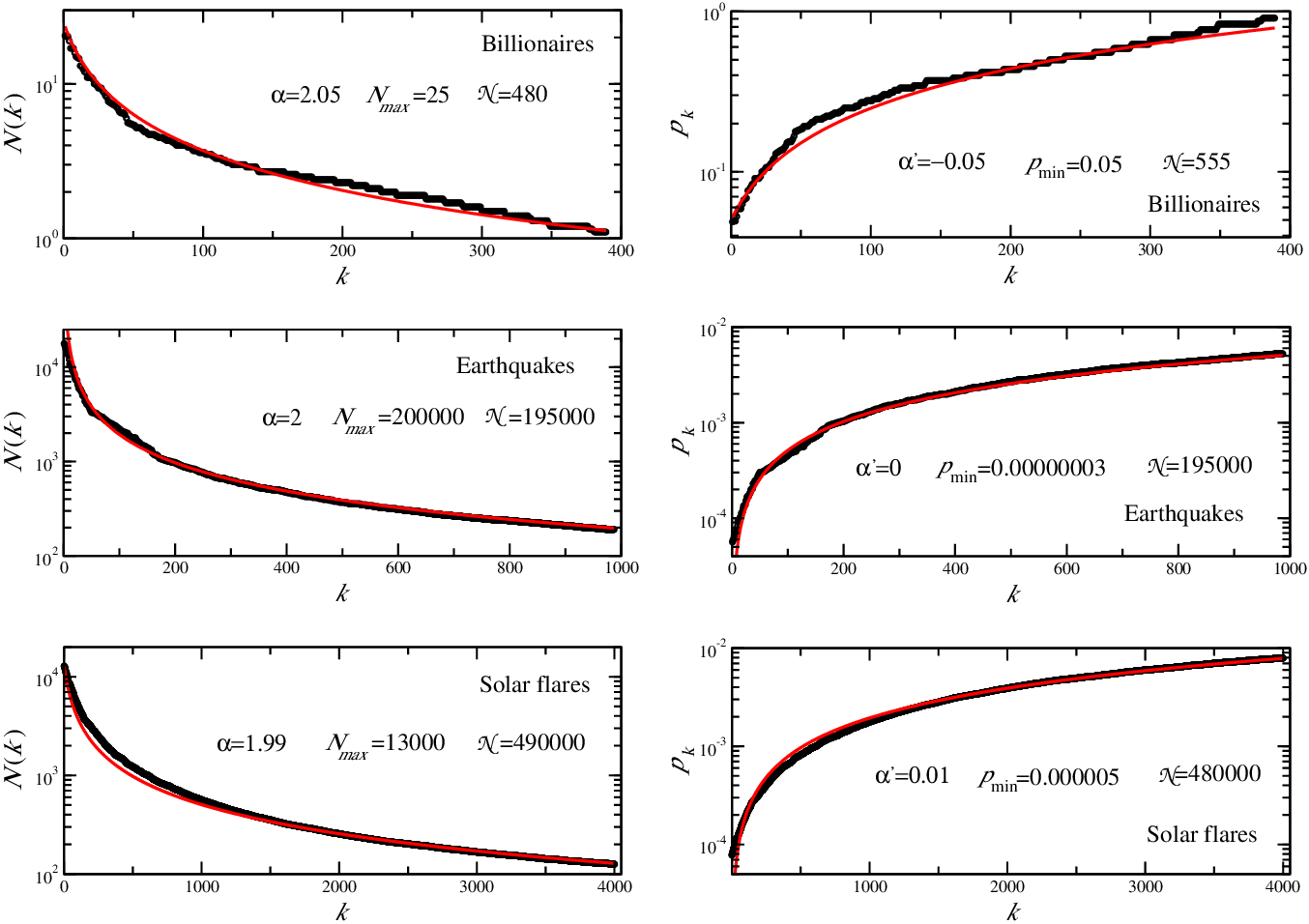}}
\caption{The same three examples in Fig. 1 are fitted with the expressions
in Eqs. (8) and (25). As it can be seen in the figure the values of $\protect%
\alpha$ needed for fitting are close to $\protect\alpha$ $\simeq$ 2 and $\protect%
\alpha ^{\prime }=2-\protect\alpha$ $\simeq$ 0. The value $\protect\alpha=2$
gives the classical Zipf law exponent, whereas the value $\alpha ^{\prime }=0$
indicates extreme configuration-space contraction. See text for description.}
\label{afoto2}
\end{center}
\end{figure*}

\begin{figure*}[ht]
\begin{center}
\centerline{\includegraphics[width=.7\textwidth]{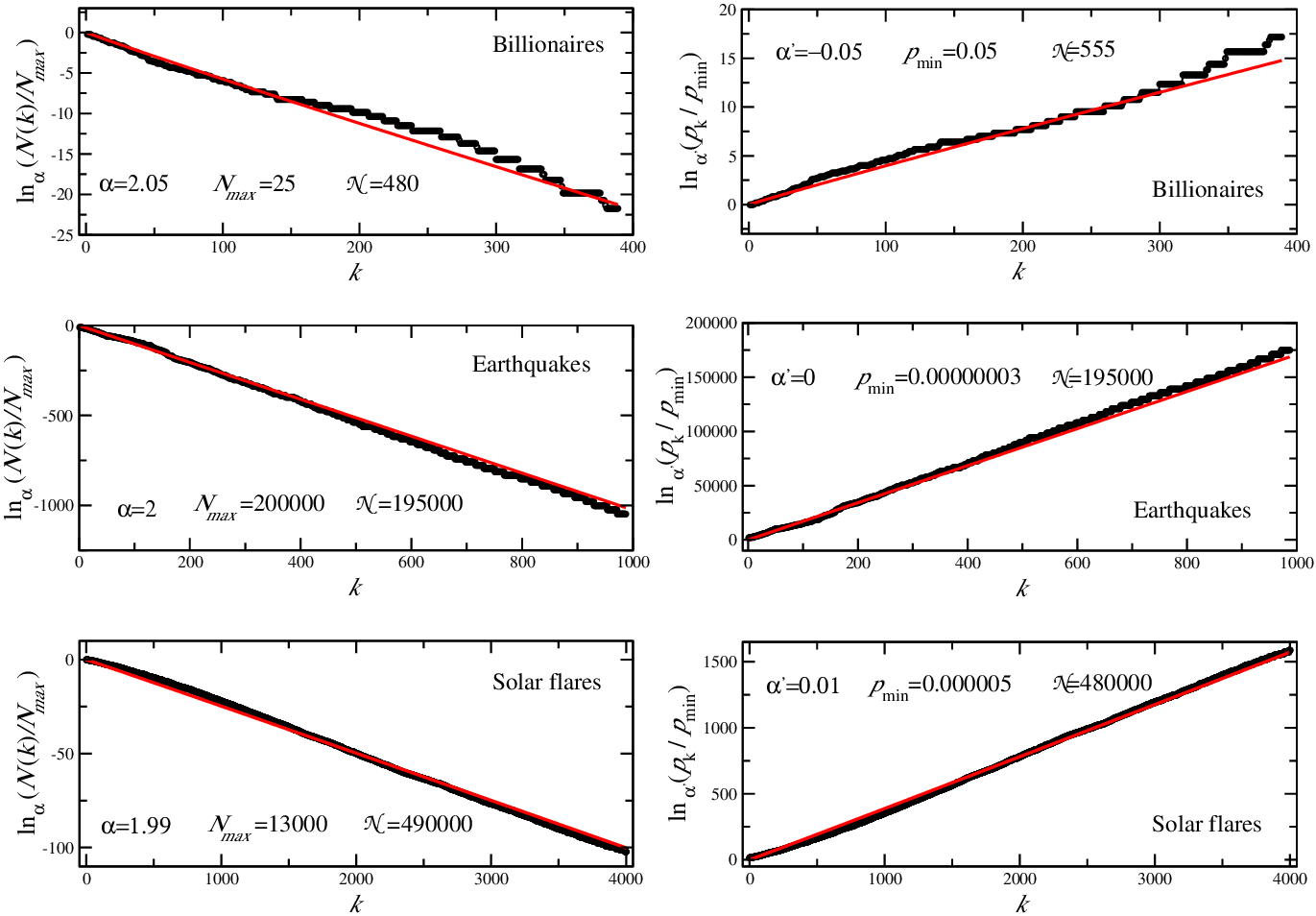}}
\caption{The same three examples in Figs. 1 and 2 plotted in $\ln_\protect%
\alpha (N(k)/N_{\max })$ (left) and $\ln_{\protect\alpha^{\prime}
}(p_k/p_{\min } )$ (right) scales. Data plotted in these scales are designed to display
linear behavior if the theoretical expressions in Eqs. (8) and (25) are
fulfilled by the data. }
\label{afoto3}
\end{center}
\end{figure*}


\begin{thebibliography}{10}

\bibitem{general1} See J.G. van der Galien (2003) in http://en.wikipedia.org/wiki/Zipfs\_law

\bibitem{schroeder} Schroeder M. (1991) Fractals, Chaos, Power Laws: Minutes from an Infinite Paradise, W.H.
Freeman and Company.

\bibitem{gell-mann}  Gell-Mann M. (1994) The Quark and the Jaguar: Adventures in the Simple and the Complex, W.H. Freeman and Company.

\bibitem{robledo1} Altamirano, C.; Robledo, A. (2011) Possible thermodynamic structure underlying the laws of Zipf and Benford Eur. Phys. J.B., 81(3):345-351. 

\bibitem{robledo2} Robledo, A. (2011) Laws of Zipf and Benford, intermittency, and critical fluctuations, Chinese Sci. Bull., 56(34):3645-3648.

\bibitem{tsallis1} Tsallis, C. (2009)  Introduction to Nonextensive Statistical Mechanics: Approaching a Complex World (Springer: New York, NY).

\bibitem{forbes1} http://www.forbes.com/billionaires. Accessed May 29, 2014.

\bibitem{earthquakes1} Southern California Earthquake Data Center,
http://www.data.scec.org. Accessed May 29, 2014.

\bibitem{solarflares1} http://tuvalu.santafe.edu/~aaronc/powerlaws/data.htm. Accessed May 29, 2014.

\bibitem{pietronero1} Pietronero L., Tosatti E., Tosatti V., Vespignani A. (2001) Explaining the uneven distribution of numbers in nature: the laws of Benford and Zipf,  Physica A, 293(1-2):297-304. 

\bibitem{thurner hanel gell-mann1} Hanel, R; Thurner, S; Gell-Mann, M. (2011) Generalized entropies and the transformation group of superstatistics. Proc Natl Acad Sci USA 108(16):6390Ð6394.

\bibitem{thurner hanel gell-mann2} Hanel, R; Thurner, S; Gell-Mann, M. (2012) Generalized entropies and logarithms and their duality relations. Proc Natl Acad Sci USA 109(47):19151-19154.

\bibitem{thurner hanel gell-mann3} Hanel, R; Thurner, S; Gell-Mann, M. (2014) How multiplicity determines entropy and the derivation of  the maximum entropy principle for complex systems.  Proc Natl Acad Sci USA 111(19):6905-6910.

\bibitem{abe1} Abe, S. (2000) Axioms and uniqueness theorem for Tsallis entropy Phys. Lett A  271(1-2):74-79.

\bibitem{robledo3} Baldovin, F. and Robledo, A. (2004), Nonextensive Pesin identity: Exact renormalization group analytical results for the dynamics at the edge of chaos of the logistic map, Phys. Rev. E  69(4 Pt 2):045202.

\bibitem{robledo4} Robledo, A. (2013) Generalized Statistical Mechanics at the Onset of Chaos, Entropy, 15(12):5178-5222.

\end{thebibliography}
\end{document}